\documentclass[]{aastex631}
\usepackage{amsmath}
\usepackage{amssymb}
\usepackage{graphicx}
\usepackage{xcolor}
\usepackage{widetable}
\usepackage{tabularx}
\usepackage{booktabs}
\usepackage{epsf}
\usepackage{longtable}

\newcommand{\cm}{cm$^{-1}$}
\newcommand{\alumina}{Al$_2$O$_3$}
\newcommand{\kcal}{kcal mol$^{-1}$}
\newcommand{\kmol}{km mol$^{-1}$}
\newcommand{\alane}{AlH$_3$}
\newcommand{\ammonia}{NH$_3$}

\begin{document}

\title{Reaction Pathway and Rovibrational Analysis of Aluminum Nitride Species\\
as Potential Dust Grain Nucleation Agents}

\author[0000-0001-6432-5638]{C. Zachary Palmer}
\affiliation{Department of Chemistry \& Biochemistry, University of Mississippi,\\
University, Mississippi, 38677, United States}

\author[0000-0003-4716-8225]{Ryan C. Fortenberry}
\affiliation{Department of Chemistry \& Biochemistry, University of Mississippi,\\
University, Mississippi, 38677, United States}

\date{\today}

\begin{abstract}

A dust nucleating agent may be present in interstellar or circumstellar media that has gone seemingly undetected and unstudied for decades. Some analyses of the Murchison CM2 meteorite suggest that at least some of the aluminum present within condensed as aluminum nitrides instead of the long studied, but heretofore undetected suite of aluminum oxides. The present theoretical study utilizes explicitly correlated coupled cluster theory and density functional theory to provide a pathway of formation from alane (AlH$_3$) and ammonia to the cyclic structure, Al$_2$N$_2$H$_4$ which has the proper Al/N ratio expected of bulk aluminum nitrides. Novel rovibrational spectroscopic constants are computed for alane and the first two formed structures, AlNH$_6$ and AlNH$_4$, along the reaction pathway for use as reference in possible laboratory or observational studies. The $\nu_8$ bending frequency for AlNH$_6$ at 755.7 \cm\ (13.23 $\mu$m) presents a vibrational transition intensity of 515 \kmol, slightly more intense than the anti-symmetric C$-$O stretch of carbon dioxide, and contains a dipole moment of 5.40 D, which is $\sim 3 \times$ larger than that of water. Thus, the present reaction pathway and rovibrational spectroscopic analysis may potentially assist in the astrophysical detection of novel, inorganic species which may be indicative of larger dust grain nucleation.

\end{abstract}


\section{Introduction} \label{sec:intro}
 Previous study \citep{Zinner91} of the Murchison CM2 chondritic meteorite seems
 to suggest the aluminum present within condensed as a form of
 aluminum nitride rather than the more commonly assumed aluminum oxides. 
 However, aluminum nitride clusters must compete
 with these aluminum oxides during dust grain nucleation and further
 formation. Alumina (\alumina) is a common aluminum oxide that
 has been theorized as a major contributor to both bulk aluminum
 oxides and their use in dust grain nucleation and formation in the
 interstellar medium (ISM) and in circumstellar media
 (CSM) \citep{Gail98,Gail99,Gobrecht16}. One of the most commonly
 attributed spectral features for bulk alumina's Al$-$O vibrational
 stretching and bending motions is seen at 13
 $\mu$m \citep{Gobrecht22,Sloan03}. Additionally, previous studies \citep{Gail99,Sloan03}
 attribute emission features at 11, 20, 28, and 32 $\mu$m to the same
 carrier as the 13 $\mu$m feature. Despite this, monomeric forms of aluminum 
 oxides like \alumina\ have yet to be detected as their formation seems to be 
 unfavorable due to many pathways hindered by endothermicities
 \citep{Chang98,Patzer05,Gobrecht22}. Further, bulk-phase aluminum oxides exhibit 
 high condensation temperatures that lead to reaction timescales that are too 
 rapid to adequetly detect such clusters \citep{Lodders03}. The presence of 
 \alumina\ in chondritic meteorite studies suggests its 
 existence in these stellar environments; without astronomical detection, 
 though, their potential role in dust grain nucleation to formation cannot be confirmed 
 \citep{Nittler08,Hutcheon94}. Thus, an investigation into an alternative 
 Al-containing species that may not condense on timescales as quickly as \alumina\ 
 is warranted to elucidate long-sought-after dust formation pathways containing 
 Al that have eluded astrochemists and astrophysicists to date.

 While Al condensates show high refractory character \citep{Lodders03}, 
 not all of its forms, in CSM, condense into large bulk solids early in stellar
 evolution and evade detection. Within the circumstellar envelope of
 IRC+10216, amongst other oxygen-rich sources, two Al-halide species have been 
 observed in the form of AlCl and AlF \citep{Cernicharo87Metals,Ziurys94F, Decin17, Saberi22}. 
 These Al-halides have also been observed within circumstellar envelopes with C/O $\sim$ 1 
 \citep{Danilovich21}. While these have
 been observed closer to the star, upon their ejection to the cooler
 regions of the stellar envelope, they are thought to deplete onto
 dust grains eluding any further detection. However, this is not the
 case for AlNC, which has been detected in larger concentrations
 within the cooler regions of the envelope and under the condensation
 temperature of the elusive \alumina\ molecule and other aluminum
 oxide clusters \citep{Ziurys02}. The presence of the Al-halides and
 AlNC suggests that Al-containing species are not solely locked up
 into aluminum oxide clusters or grains. Thus, O-deficient,
 Al-containing species can be formed within the warmer regions of the
 stellar envelope and persist long enough in the cooler, outer
 circumstellar dust clouds before depositing or aggregating and
 disappearing from rovibrational detection.

 Additionally, a radioisotope of Al, $^{26}$Al, was identified in its
 fluorinated form, $^{26}$AlF, near the stellar merger of CK
 Vulpus \citep{Kaminski18}. $^{26}$AlF is believed to have been introduced into the
 surrounding stellar region during the collision of the binary system
 yet persisted on a timescale long enough for
 detection \citep{Kaminski15a}. Much like the elusive \alumina, both the
 stable $^{27}$Al and an extinct form of $^{26}$Al, which later
 decayed into the $^{26}$Mg isotope based on isotopic abundance studies,
 were found within the Murchison CM2 chondritic meteorite. An
 investigation of the ratio of $^{26}$Al/$^{27}$Al suggests that the
 $^{26}$Al present is in larger abundance in at least this meteorite
 compared to the ratio in the solar system at large. In any case, the 
 presence of an aluminum nitride system within Murchison CM2 is suggested 
 \textit{via} a comparison of the SiC abundance and CN$^{-}$/C$^{-}$ 
 ratio and the Al and N present within. Such a correlation suggests 
 that the Al in the meteorite condensed not as some form of aluminum 
 oxide, like \alumina, but rather as some form of aluminum nitride 
 system \citep{Zinner91}. As discussed previously, the Al$-$N moiety 
 is not unheard of in CSM from the presence of AlNC. Previous computational 
 studies have calculated an Al$-$N bond strength of $-$105.0 \kcal, which is
 stronger than the bond strength of the N$-$C bond at $-$77.8
 \kcal\ \citep{Doerksen20}. This larger bond strength of the 
 Al$-$N bond, relative to the N$-$C bond in AlNC, suggests the stability 
 of Al$-$N bonds in CSM and the ISM. 
 
 The study of the Murchison CM2 meteorite implies that the aluminum
 nitride systems present within play a key role in the nucleation of
 the graphite that composes the bulk of the carbonaceous material
 within the meteor by catalyzing the graphite’s inhibited
 nucleation \citep{Czyzak82,Nuth85,Zinner91}. With that, these new
 Al-containing species may be instrumental for the nucleation of
 aggregated material onto other solar system bodies, such as comets,
 asteroids, and other meteors. That being said, these conclusions
 justify an investigation into the presence of unreported aluminum
 nitride molecular systems in CSM and the ISM that may potentially
 assist in elucidating the processes of how Al-bearing species get
 from the gas phase into their bulk solid-phase dust grain
 counterparts.

 As aluminum nitride systems have not been characterized in any
 stellar environment, no gas-phase observational or experimental
 spectroscopic data are available to begin the search for such
 species. Thus, the present quantum chemical study
 provides reference data for a proposed pathway of formation from the 
 aluminum hydride, alane, (\alane) and ammonia (\ammonia) into the first cyclic
 species along said pathway. Regardless of the \alane\ molecule's lack of 
 interstellar detection, the present study utilizes \alane\ as the main 
 source of aluminum given its simplicity as a metal hydride and its closed-shell 
 configuration. The \alane\ molecule is used in place of the more simple, 
 and previously detected \citep{Kaminski16}, aluminum mono-hydride (AlH) as 
 previous computational studies suggest that access to more hydrogen atoms 
 assists the progress of the reaction pathway \citep{Grosselin22,Flint23b}. 
 The rovibrational spectroscopic data herein will be instrumental in supporting 
 the currently available spectroscopic telescopes and observational technologies like the
 \textit{James Webb Space Telescope} (JWST) for its efficiency in
 probing the near- to mid-IR spectrum with its Near-Infrared Spectrograph 
 (NIRSpec) and Mid-Infrared Instrument (MIRI)
 instruments. Where available, the provided rotational
 data will assist microwave spectroscopic observatories like the
 \textit{Atacama Large Millimeter/sub-millimeter Array} (ALMA). To
 that end, the rovibrational spectroscopic analysis provided should 
 aid in the potential astrophysical identification of aluminum
 nitride species that have yet to be characterized and may
 contribute to the dust grain nucleation and formation processes
 present in CSM and the ISM.
 
 \section{Computational Methods}
 \subsection{Reaction Mechanism Methods}
 Unless otherwise stated, all geometry optimizations, single-point
 energy (SPE) computations, and zero-point vibrational energy
 corrections for the reactants, intermediates, and products of the
 aluminum nitride reaction pathway are conducted utilizing coupled
 cluster theory at the singles, doubles, and perturbative triples
 level [CCSD(T)] \citep{Rag89, Shavitt09, ccreview}. For an additional
 gain in accuracy, the CCSD(T) level of theory is corrected within the
 explicitly correlated F12b formalism \citep{Adler07,Knizia09} along
 with its corresponding cc-pVTZ-F12 basis
 set \citep{Peterson08,Yousaf08}. The aforementioned level of theory
 will henceforth be abbreviated as ``F12-TZ.'' Geometry optimization
 and harmonic frequency computations for all transition states along
 the pathway are conducted with the B3LYP density
 functional \citep{LYP86,LYP88} along with the aug-cc-pVTZ correlation
 consistent Dunning basis set \citep{aug-cc-pVXZ}. After the transition
 states are optimized at the B3LYP/aug-cc-pVTZ level of theory,
 single-point energy computations are computed at the F12-TZ
 level \citep{Olmedo21}, and are subsequently corrected with
 B3LYP/aug-cc-pVTZ zero-point vibrational energies (ZPVEs). All minima
 along the reaction pathway are computed \textit{via} the MOLPRO
 2022.2 suite of quantum chemical packages \citep{MOLPRO}. Finally, all
 transition state computations on the pathway are conducted
 \textit{via} GAUSSIAN16 \citep{g16}.

 \subsection{Rovibrational Spectroscopic Methods}
 The highly accurate rovibrational spectroscopic constants are
 computed utilizing the quartic force field (QFF) approach. A QFF is a
 fourth-order Taylor series expansion of the potential energy portion
 of the internuclear Watson Hamiltonian \citep{Fortenberry19QFF}. QFFs
 as employed herein have been able to produce rovibrational
 spectroscopic constants within 1\% of experimental values for many
 molecular systems \citep{Huang08, Huang09, Huang11, Zhao14,
   Huang13NNOH+, Fortenberry14C2H3+, Fortenberry15SiC2, Kitchens16,
   Fortenberry17IJQC, Fortenberry18C3H2, Gardner21}. F12-TZ has been shown to
 produce accurate fundamental vibrational frequencies and rotational
 constants at far less computational cost compared to other QFF
 methods \citep{Martin14,Agbaglo19a,Agbaglo19c,Palmer20,Palmer22a}.
 Additionally, the QFF implemented in this work is conducted within
 the recently developed automated PBQFF
 framework \citep{Westbrook23_pbqff}. The PBQFF procedure begins with a
 geometry optimzation utilizing MOLPRO \citep{MOLPRO} at the F12-TZ
 level of theory with tight convergence criteria. From there, the
 optimized Cartesian geometry is displaced by 0.005 \AA\,
 respective of bond lengths or angles, to mimic a QFF but is truncated
 to the second-order yielding a Cartesian harmonic force field. 
 SPE computations for every displacement
 are computed and used to generate a harmonic force constant (FC)
 matrix. The normal coordinates are then extracted from the resulting 
 mass-weighted Hessian matrix for the given molecular species. The 
 optimized molecular geometry is then displaced along these normal coordinates 
 to compute the rest of the semi-diagonal QFF, and SPE computations are performed 
 at each normal coordinate displacement.

 Once the SPE computations are finished, the final normal FCs are
 computed directly utlizing a finite differences procedure. The
 normal coordinate FCs are then passed to a second-order vibrational and
 rotational perturbation theory (VPT2) \citep{Mills72, Watson77, Papousek82, Franke21} 
 algorithm within the PBQFF framework itself. From this, the harmonic 
 frequencies, fundamental anharmonic vibrational frequencies, 
 vibrationally-averaged rotational constants, singly-vibrationally-excited 
 rotational constants, quartic distortion constants, and sextic distortion 
 constants are produced. Additionally, if present in the analysis, the type-1 and -2
 Fermi resonances and Coriolis resonances are taken into account as it has 
 been shown to increase the accuracy of the computed rovibrational spectroscopic
 constants \citep{Martin97,Martin95}. To further assist in potential
 astrophysical detection, dipole moments for AlNH$_6$ and AlNH$_4$ are
 computed at the F12-TZ level within MOLPRO 2022.2 \citep{MOLPRO}.
 Finally, double-harmonic infrared intensities computed using
 GAUSSIAN16 \citep{g16} at the MP2/cc-pVDZ \citep{MP2,Dunning89} level of theory 
 are included to assist in the detection of these species in the infrared (IR). 
 Computed intensities at this level of theory have been show to produce 
 semi-quantitative agreement with higher levels of theory for far less 
 computational costs \citep{Yu15NNHNN+,Finney16,Westbrook21a}. 
 
 In addition to IR intensities, absorption cross sections, $\sigma$, are 
 given for all applicable vibrational frequencies. The absorption cross sections are 
 derived utilizing the formula in EQ.~\ref{abscs}, where "N$_a$" is Avogadro's number, 
 and "$\epsilon$" is the molar absorption coefficient.
 
\begin{equation} \label{abscs}
     \sigma = \frac{ln(10)\times10^3}{N_a}\times\epsilon
 \end{equation}
 
 The molar absorption coefficients for each vibrational frequency are 
 computed following the formula in EQ.~\ref{epsmax} \citep{IReps}, where "I$_{IR}$" is 
 the IR intensity computed above and "w" is the resolving power of the given observing 
 telescope.

 \begin{equation} \label{epsmax}
     \epsilon = 27.648\times\frac{I_{IR}}{w}
 \end{equation}
 
The resolution, "R," of the NIRSpec and MIRI on the \textit{JWST}, 
at their respective operating wavelengths, are provided and are trivially 
converted to the resolving power in EQ.~\ref{res} \citep{MIRI21,NIRSpec22}.

 \begin{equation} \label{res}
     R = \frac{\lambda}{\Delta\lambda}
 \end{equation}

 \section{Results and Discussion}
 \subsection{Reaction Pathway Analysis}
 
 \begin{figure*} 
   \centering
   \includegraphics[width=\textwidth]{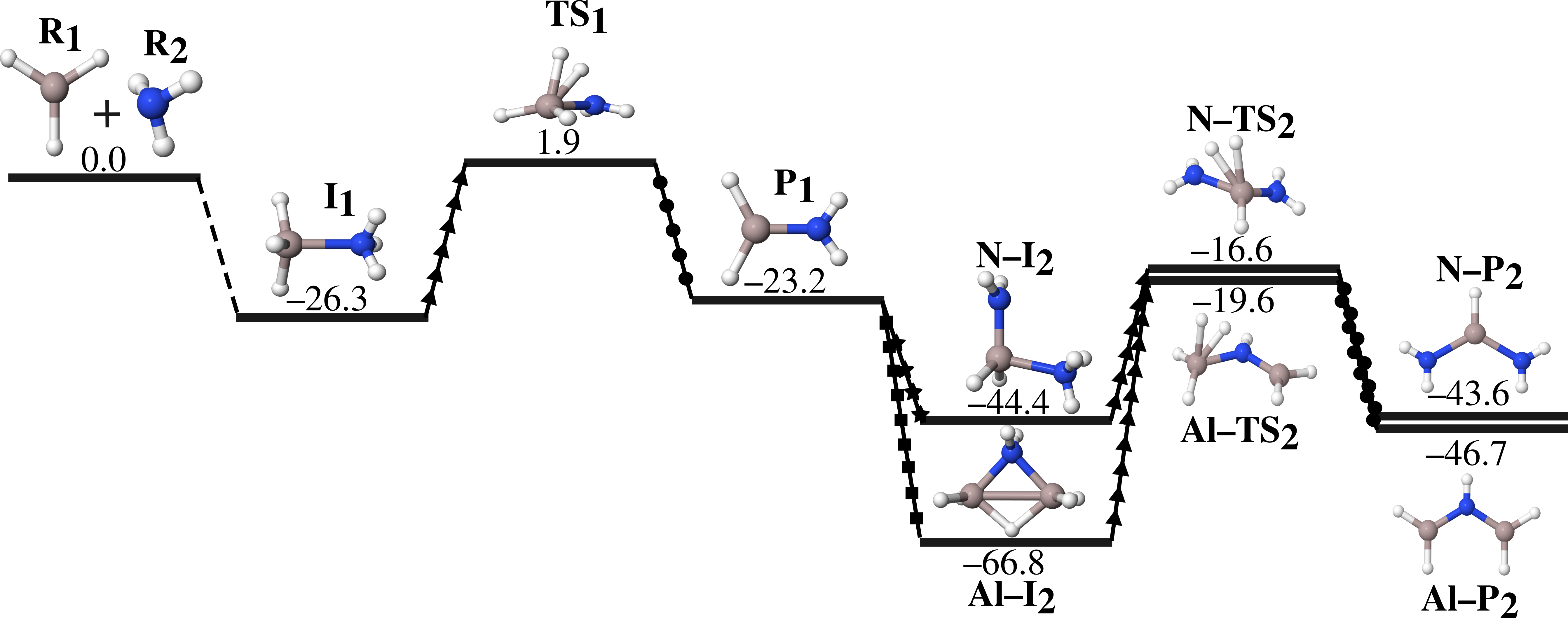}
   \caption{Reaction coordinate diagram from \alane\ and \ammonia\ to
     N$-$ and Al$-$P$_2$. The arrowed lines indicate isomerization,
     the circular lines indicate H$_2$ departure, the starred lines indicate
     \ammonia\ addition, and the squared lines indicate \alane\
     addition. Relative energies are in \kcal. White atoms indicate H, 
     blue atoms indicate nitrogen, and grey beige indicate Al.}
 \label{react1}
 \end{figure*}

 \begin{figure*} 
   \centering
   \includegraphics[width=\textwidth]{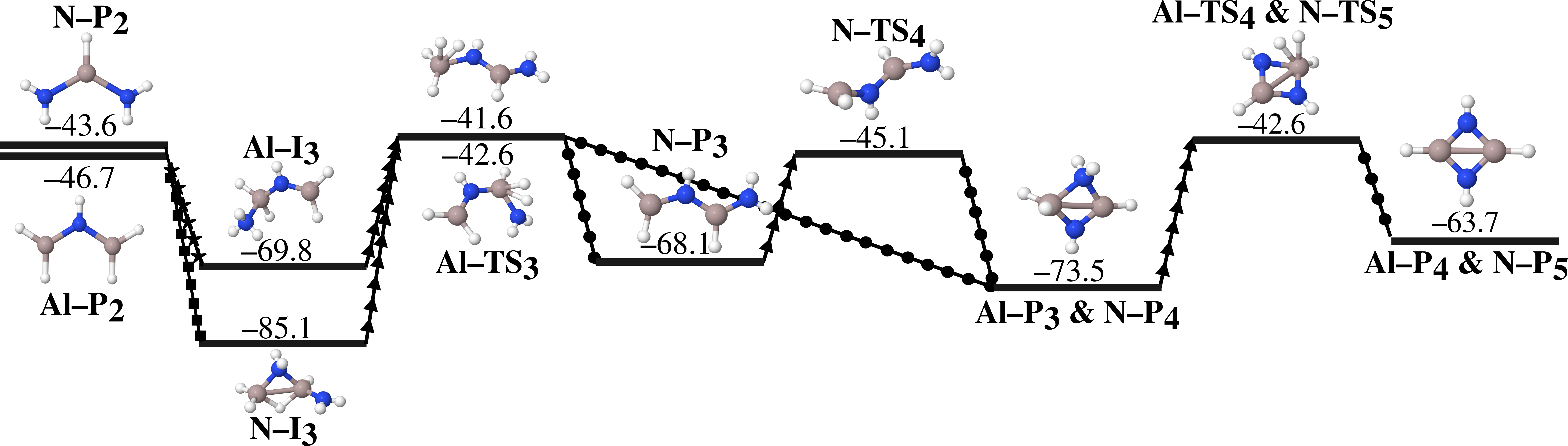}
   \caption{The continued reaction coordinate diagram from N$-$ and
     Al$-$P$_2$ to cyclic-Al$_2$N$_2$H$_4$. The arrowed lines indicate isomerization,
     the circular lines indicate H$_2$ departure, the starred lines indicate
     \ammonia\ addition, and the squared lines indicate \alane\
     addition. Relative energies are in \kcal. White atoms indicate H, 
     blue atoms indicate nitrogen, and grey beige indicate Al.}
 \label{react2}
 \end{figure*}

 \begin{table}
     \centering
     \caption{Symbol definitions and chemical formulae for the species in the present reaction pathway.}
     \begin{tabular}{l|l|c}
          Symbol & Definition & Chemical Formula \\
          \midrule
          R$_1$       & Reactant 1 & AlH$_3$ \\
          R$_2$       & Reactant 2 & NH$_3$ \\
          I$_1$       & Intermediate 1 & AlNH$_6$ \\
          TS$_1$      & Transition State 1 & AlNH$_6$ \\
          P$_1$       & Product 1 & AlNH$_4$ \\
          N$-$I$_2$   & Nitrogen Intermediate 2 & AlN$_2$H$_7$ \\
          Al$-$I$_2$  & Aluminum Intermediate 2  & Al$_2$NH$_7$ \\
          N$-$TS$_2$  & Nitrogen Transition State 2 & AlN$_2$H$_7$ \\
          Al$-$TS$_2$ & Aluminum Transition State 2 & Al$_2$NH$_7$ \\
          N$-$P$_2$   & Nitrogen Product 2 & AlN$_2$H$_5$ \\
          Al$-$P$_2$  & Aluminum Product 2 & Al$_2$NH$_5$ \\
          N$-$I$_3$   & Nitrogen Intermediate 3 & Al$_2$N$_2$H$_8$ \\
          Al$-$I$_3$  & Aluminum Intermediate 3 & Al$_2$N$_2$H$_8$ \\
          N$-$TS$_3$  & Nitrogen Transition State 3 & Al$_2$N$_2$H$_8$ \\
          Al$-$TS$_3$ & Aluminum Transition State 3 & Al$_2$N$_2$H$_8$ \\
          N$-$P$_3$   & Nitrogen Product 3          & Al$_2$N$_2$H$_6$ \\
          N$-$TS$_4$  & Nitrogen Transition State 4 & Al$_2$N$_2$H$_6$ \\
          Al$-$P$_3$  & Aluminum Product 3          & Al$_2$N$_2$H$_6$ \\
          N$-$P$_4$   & Nitrogen Product 4          & Al$_2$N$_2$H$_6$ \\
          Al$-$TS$_4$ & Aluminum Transition State 4 & Al$_2$N$_2$H$_6$ \\
          N$-$TS$_5$  & Nitrogen Transition State 5 & Al$_2$N$_2$H$_6$ \\
          Al$-$P$_4$  & Aluminum Product 4          & Al$_2$N$_2$H$_4$ \\
          N$-$P$_5$   & Nitrogen Product 5          & Al$_2$N$_2$H$_4$ \\
     \end{tabular}
     \label{symlabs}
 \end{table}
 
 The reaction coordinate profile shown in Figs.~\ref{react1} and \ref{react2} along 
 with their corresponding equilibrium geometries, and definitions and chemical formulae in 
 Table \ref{symlabs}, shows the continual addition of
 equivalents of \alane\ and \ammonia. Additions of \alane\ and \ammonia\ 
 produce progressively larger datively-bonded structures but contain an 
 initially raised transition barrier from I$_1$ to the 
 P$_1$. In order for the reaction to progress to larger aluminum 
 nitride systems, I$_1$ must overcome the aforementioned barriered TS$_1$ that sits 
 1.9 \kcal\ higher than the reactants and lose an H$_2$ molecule to form P$_1$. This is in stark 
 contrast with previous computed reaction
 schemes \citep{Grosselin22, Gobrecht22} on \alane\ and water that
 produce aluminum oxides with submerged barriers of formation. These
 barrierless formation most likely govern the short timescales of
 leading to the underdetection of a suitable carrier of the Al$-$O
 features, like \alumina\, in CSM or the ISM. Regardless, overcoming this
 barrier requires sufficiently high temperatures, $\sim$960
 K, in the aluminum nitride species' formation environment from the
 surrounding stellar environments to begin forming larger clusters.

 Obviously, the cold, diffuse ISM at $\sim$10$-$40 K is not a likely
 region for the formation of P$_1$ to occur. However, as stated
 earlier, the previous work on Murchison CM2 \citep{Zinner91} suggests
 that the theorized aluminum nitride systems present in the meteorite
 may have nucleated the meteorite's dust grain formation. If this is
 the case, then the aluminum nitride system would have been formed in
 some region warm enough for its own production and for dust grain
 formation to occur. The warm environments of inner protoplanetary
 disks can achieve a range of temperatures anywhere from 500$-$1500
 K \citep{Boss98} and contain ample material for molecular synthesis to
 begin that may potentially lead to seeding and nucleating processes
 of dust grain formation. Additionally, highly evolved asymptotic
 giant branch (AGB) stars are known to exhibit much higher
 temperatures within 1$-$2 stellar radii at 2000$-$3000 K compared to
 the warm inner protoplanetary disks \citep{Maercker22}. AGB stars also
 contribute considerable dust forming material to their surrounding
 environment, and the ISM at large, given the stars substantial mass-loss
 rates \citep{Hofner18,Ferrarotti06}. If a region of CSM or 
 the ISM contains sufficient temperatures and has the necessary material, 
 the proper conditions toward formation the aluminum nitride species along the 
 reaction pathway proposed herein will be satisfied. Once these aluminum nitrides 
 have been formed, they may likely go on to contribute to dust grain nucleation and 
 formation as suggested in the study of the Murchison CM2 meteorite.

 Once the transitional barrier, shown in Fig.~\ref{react1}, is overcome
 there will be enough latent energy within the system to further
 progress through the now relatively submerged pathway of
 formation. Thus, the reaction pathway is no longer nearly as limited by the
 surrounding temperature and can further condense into larger and
 larger dust grains in both the warmer and cooler regions of the CSM
 or ISM. In any case, from the P$_1$ structure, additions of \alane\ or
 \ammonia\ lead to either the N$-$I$_2$ or the Al$-$I$_2$ structure. The
 addition of the \alane\ leads to a lower energy cyclic structure with
 a three-center two-electron bond commonly seen in trivalent species
 containing an empty \textit{p}-orbital like aluminum and boron \citep{Mayer89}. Upon
 the loss of the H$_2$ molecule, seen in the N/Al$-$TS$_2$ species,
 both pathways lead to their respective products (P$_2$) of similar
 structure. Interestingly, if an equivalent of \alane\ is added to
 N$-$P$_2$, it stabilizes into N$-$I$_3$, which is structurally similar
 to Al-I$_2$ with a three-center two-electron bond. The extra stability 
 of the three-center two-electron bond in N$-$I$_3$ yields the lowest 
 energy structure of the present reaction pathway. While N$-$I$_3$ is 
 the lowest energy structure, the reaction pathway should still progress 
 further not only due to the energy from the ambient temperature of its 
 formation environments but also from collision with other molecular 
 citizens present within the same regions.

 Further departure of an H$_2$ molecule from (N or Al)$-$I$_3$ yields the cyclic 
 Al$-$P$_3$ structure, whereas N$-$P$_3$ is an open chain structure. 
 The open-chain N$-$P$_3$ structure must overcome a torsional barrier 
 before it becomes the cyclic N$-$P$_4$. At this point along the pathway, 
 the paths of additions of \alane\ and \ammonia\ converge to the same 
 structure. After a final loss of and H$_2$ molecule in Al$-$TS$_4$ and N$-$TS$_5$,
 the proposed reaction pathway concludes with the final cyclic structure of
 Al$-$P$_4$ and N$-$P$_5$ (Al$_2$N$_2$H$_4$). This Al$_2$N$_2$H$_4$ species 
 may then go on to form larger aluminum nitride clusters that potentially 
 contribute to the nucleation and formation of dust grains in the universe. 
 Given that the above reaction pathway contains no carbon- or oxygen-containing 
 species, the present reaction pathway may contribute to the aluminum dust grain 
 nucleation pathways in regions where the C/O ratio is $\sim 1$. However, in 
 oxygen-rich environments it may compete with the aluminum-oxygen pathways 
 described in previous studies \citep{Gail98,Gail99,Gobrecht22}. This may not be 
 the case for carbon-dominated environments, C/O $\geq$ 1, given the lack of oxygen 
 and the aluminum-carbon motif has yet to be characterized in CSM or the ISM in any 
 form other than AlNC. Therefore, the reaction pathway investigated in this work may 
 assist in the potential detection of aluminum-containing dust grain nucleation species 
 that that have gone undetected, and understudied, in the literature for decades.
 
\begin{center}
\begin{longtable}{|lr|c|c|c|}
\caption{Rotational Constants for AlH$_3$, AlNH$_4$, and AlNH$_6$} \label{rot} \\
\hline
\multicolumn{1}{|l}{Const.} & \multicolumn{1}{r|}{Units} & \multicolumn{1}{c|}{AlH$_3$} & \multicolumn{1}{c|}{AlNH$_4$} & \multicolumn{1}{c|}{AlNH$_6$} \\
\endfirsthead

\multicolumn{5}{c}%
{{\tablename\ \thetable{} -- continued from previous page}} \\
\hline \multicolumn{1}{|l}{Const.} & \multicolumn{1}{r|}{Units} & \multicolumn{1}{c|}{AlH$_3$} & \multicolumn{1}{c|}{AlNH$_4$} & \multicolumn{1}{c|}{AlNH$_6$} \\ \hline
\endhead

\hline \multicolumn{5}{|r|}{{Continued on next page}} \\ \hline
\endfoot

\hline
\endlastfoot

\hline                                
$A_{e}$  & MHz   & 133833.5 & 95478.0  & 49434.0  \\
$B_{e}$  & MHz   & 133833.5 & 13074.1  & 9048.5   \\
$C_{e}$  & MHz   &  66916.7 & 11499.5  & 9048.4   \\
$A_{0}$  & MHz   & 132407.6 & 95063.3  & 49154.6  \\
$B_{0}$  & MHz   & 132407.6 & 13015.9  & 8916.2   \\
$C_{0}$  & MHz   &  66034.4 & 11430.3  & 8916.4   \\
$A_{1}$  & MHz   & 131376.9 & 94810.8  & 49090.3  \\
$B_{1}$  & MHz   & 131376.9 & 12994.8  & 8913.7   \\
$C_{1}$  & MHz   &  65518.7 & 11412.4  & 8913.9   \\
$A_{2}$  & MHz   & 131652.0 & 94636.1  & 49025.6  \\
$B_{2}$  & MHz   & 131652.0 & 12999.7  & 8914.2   \\
$C_{2}$  & MHz   &  66346.6 & 11411.6  & 8914.4   \\
$A_{3}$  & MHz   & 131457.8 & 94432.1  & 48865.1  \\
$B_{3}$  & MHz   & 131457.8 & 13009.1  & 8920.6   \\
$C_{3}$  & MHz   &  65043.4 & 11418.4  & 8920.8   \\
$A_{4}$  & MHz   & 132824.6 & 94214.4  & 48926.0  \\
$B_{4}$  & MHz   & 132824.6 & 13012.7  & 8922.5   \\
$C_{4}$  & MHz   &  66244.7 & 11415.5  & 8922.7   \\
$A_{5}$  & MHz   &          & 95436.9  & 49035.0  \\
$B_{5}$  & MHz   &          & 13034.2  & 8927.4   \\
$C_{5}$  & MHz   &          & 11420.7  & 8927.6   \\
$A_{6}$  & MHz   &          & 95225.2  & 49144.2  \\
$B_{6}$  & MHz   &          & 12981.5  & 8881.5   \\
$C_{6}$  & MHz   &          & 11369.6  & 8881.7   \\
$A_{7}$  & MHz   &          & 95582.8  & 49064.1  \\
$B_{7}$  & MHz   &          & 13017.0  & 8918.3   \\
$C_{7}$  & MHz   &          & 11441.5  & 8918.5   \\
$A_{8}$  & MHz   &          & 97565.1  & 49309.6  \\
$B_{8}$  & MHz   &          & 13052.9  & 8904.7   \\
$C_{8}$  & MHz   &          & 11398.1  & 8904.9   \\
$A_{9}$  & MHz   &          & 93906.6  & 49191.6  \\
$B_{9}$  & MHz   &          & 12981.2  & 8878.7   \\
$C_{9}$  & MHz   &          & 11444.4  & 8878.9   \\
$A_{10}$ & MHz   &          & 94707.4  & 49146.8  \\
$B_{10}$ & MHz   &          & 12977.0  & 8807.8   \\
$C_{10}$ & MHz   &          & 11433.0  & 8808.0   \\
$A_{11}$ & MHz   &          & 93227.5  & 49487.3  \\
$B_{11}$ & MHz   &          & 12975.5  & 8900.9   \\
$C_{11}$ & MHz   &          & 11429.1  & 8901.0   \\
$A_{12}$ & MHz   &          & 96185.7  & 49145.0  \\
$B_{12}$ & MHz   &          & 13038.4  & 8875.0   \\
$C_{12}$ & MHz   &          & 11430.9  & 8875.3   \\
$\mu$    & D     &    0.0   & 1.08     &   5.40   \\
\hline
\end{longtable}
\end{center}

 \subsection{Rovibrational Spectroscopic Analysis}

\begin{table}[b]
\centering
\caption{Quartic and sextic distortion constants in the Watson
  A-reduced Hamiltonian for AlNH$_4$ and AlNH$_6$}
\label{aham}
\begin{tabular}{lrr}
Const.        &    AlNH$_4$       &      AlNH$_6$           \\
\midrule
$\Delta_{J}$  &   12.151 (kHz) &     16.330 (kHz)     \\
$\Delta_{K}$  &    1.630 (MHz) &     76.397 (kHz)     \\
$\Delta_{JK}$ &  192.297 (kHz) &     55.756 (kHz)     \\
$\delta_{J}$  &    1.621 (kHz) &    149.664 (mHz)     \\
$\delta_{K}$  &  140.576 (kHz) &  $-$36.525 (kHz)     \\
\midrule
$\Phi_{J}$    &    2.955 (mHz) &  $-$33.349 (mHz)     \\
$\Phi_{K}$    &  122.517 (Hz)  & $-$188.055 (Hz)      \\
$\Phi_{JK}$   &    2.261 (Hz)  &  $-$79.589 (Hz)      \\
$\Phi_{KJ}$   & $-$9.071 (Hz)  &    268.269 (Hz)      \\
$\phi_{j}$    &    2.638 (mHz) &     18.738 ($\mu$Hz) \\
$\phi_{jk}$   &    1.188 (Hz)  &   $-$9.396 (Hz)      \\
$\phi_{k}$    &   81.816 (Hz)  &  $-$57.226 (MH)      \\
\end{tabular}	   
\end{table}

\begin{table}
\centering
\caption{Quartic and sextic distortion constants in the Watson
  S-reduced Hamiltonian for AlH$_3$, AlNH$_4$, and AlNH$_6$}
\label{sham}
\begin{tabular}{lrrr}
Const.        &       AlH$_3$        &      AlNH$_4$           &      AlNH$_6$          \\
\midrule
$D_{J}$       &      6.338 (MHz)  &     11.486 (kHz)     &     16.330 (kHz)    \\
$D_{JK}$      &  $-$11.275 (MHz)  &    196.288 (kHz)     &     55.756 (kHz)    \\
$D_{K}$       &      5.287 (MHz)  &      1.627 (MHz)     &     76.397 (kHz)    \\
$d_{1}$       &  $-$31.927 (Hz)   &   $-$1.621 (kHz)     & $-$149.664 (mHz)    \\
$d_{2}$       &    158.915 (Hz)   & $-$332.607 (Hz)      &      8.511 (mHz)    \\
\midrule
$H_{J}$       &    885.237 (Hz)   &   $-$3.890 (mHz)     &  $-$33.344 (mHz)    \\
$H_{JK}$      &   $-$3.292 (kHz)  &    862.837 (mHz)     &    374.006 (mHz)    \\
$H_{KJ}$      &      3.939 (kHz)  &   $-$4.307 (Hz)      &      1.725 (Hz)     \\
$H_{K}$       &   $-$1.530 (kHz)  &    119.158 (Hz)      &   $-$1.474 (Hz)     \\
$h_{1}$       &  $-$23.100 (mHz)  &      1.787 (mHz)     &     23.397 ($\mu$Hz)\\
$h_{2}$       &  $-$34.752 (mHz)  &      3.422 (mHz)     &   $-$2.245 ($\mu$Hz)\\
$h_{3}$       &  $-$52.810 (Hz)   &    851.524 ($\mu$Hz) &   $-$4.658 ($\mu$Hz)\\
\end{tabular}
\end{table}

As stated previously, some of the aluminum nitride systems investigated in
this work have little-to-no previous experimental or observational
data of any type. Therefore, the computed rovibrational spectroscopic
constants for R$_1$ (\alane), I$_1$ (AlNH$_6$), and P$_1$ (AlNH$_4$) as
shown in Fig.~\ref{react1} reported herein are reference data necessary for laboratory analysis 
and potential astrophysical detection. Presently, AlH$_3$ and AlNH$_4$ have Ar 
matrix spectroscopic data in the literature, and AlNH$_4$ and AlNH$_6$ have had 
previous theoretical studies conducted. Several of the previous theoretical 
studies \citep{Davy94, Leboeuf95, Marsh92} for AlNH$_4$ and AlNH$_6$ only 
provide an analysis of the harmonic frequencies and their structural character. 
However, The work herein provides anharmonic vibrational 
frequencies and rotational spectroscopic constants at a more rigorous level of theory.

In order to provide a full rovibrational profile for the aluminum nitride species 
investigated herein, the equilibrium, vibrationally averaged, and vibrationally 
excited rotational constants are reported in Table \ref{rot}. Also, the quartic 
and sextic distortion constants from the A- and S-reduced Watson Hamiltonians are given in
Tables \ref{aham} and \ref{sham}, respectively. Additionally, the formation pathways to 
larger aluminum nitride species assume the presence of \alane\ in the same regions. As
\alane\ has not yet been observed in CSM or the ISM, perhaps a
consequence of its rapid reaction with ammonia or water, the present
study also provides the reference data for this species for
completeness. To that end, the vibrational profile for \alane,
AlNH$_6$, and AlNH$_4$ are given in Tables \ref{alh3_freq},
\ref{alnh6_freq}, and \ref{alnh4_freq}, respectively. 

\begin{table}
\centering
\caption{Vibrational frequencies (\cm), IR Intensities (km
  mol$^{-1}$) with absorption cross sections in parentheses (10$^{-14}$ cm$^2$), and 
  wavelength ($\mu$m) for D$_{3h}$ AlH$_3$}
\label{alh3_freq}
\begin{widetabular}{\textwidth}{llcrrrrr}
  Mode      &   Symm. & Desc. & Harm.  & Anharm. & $\mathnormal{f}$ ($\sigma$) & $\lambda$ & Expt.$^a$ \\
  \midrule
  $\nu_{1}$ & $ A_1'$ & AlH Stretch           & 1951.4 & 1890.7  & 0 (0) & 5.29 &       \\
  $\nu_{2}$ & $ A_2"$ & OPB                   &  710.5 &  707.8  & 387 (200) & 14.13 & 697.8 \\
  $\nu_{3}$ & $ E'  $ & Anti-sym. AlH Stretch & 1955.7 & 1889.7  & 251 (40) & 5.29 & 1882.8\\
  $\nu_{4}$ & $ E'  $ & HAlH Anti-sym. Bend   &  798.6 &  789.1  & 234 (100) & 12.67 & 783.4 \\
  ZPT &         &   & & 4052.6 &            \\
  \midrule
\end{widetabular}
\\
\raggedright $^a$Experimental Ar matrix FTIR spectroscopy \citep{Chertihin93}.
\end{table}

As stated above, previous Ar matrix spectroscopic data \citep{Chertihin93} 
exists for \alane\ and characterizes three of the four vibrational frequencies 
available, shown in Table \ref{alh3_freq}. The out-of-plane
bending mode, $\nu_2$, differs the furthest from experiment at 10 \cm\ or 1.4\%. 
The closest to experiment
is $\nu_4$, the H$-$Al$-$H anti-symmetric bending motion,
differing by 5.7 \cm\ or 0.7\%. While the F12-TZ
anharmonic vibrational frequencies exhibit relatively large differences compared to
experiment, it should be noted that Ar matrix spectroscopy is known to
cause a shift in the true vibrational frequency \citep{Pimentel63}.
Thus, most of the discrepancy between the current high-level quantum
chemical computations and experiment should be attributed to the Ar
matrix shifts. Further, previous computational studies of
Al-containing species utilizing the F12-TZ methodology have produced
vibrational frequencies within 0.4\% of the gas-phase experimental
value \citep{Fortenberry20AlOH}. Given the higher-order, D$_{3h}$ symmetry
exhibited by \alane\ the permanent dipole of this molecule is zero,
thus rendering it undetectable \textit{via} rotational
spectroscopy. For this reason, the anharmonic vibrational frequencies
computed in this work are even more crucial for potential spaced-based IR spectroscopic 
telescopes. 

Further, \alane\ exhibits very intense vibrational transitions
with the $\nu_2$, out-of-plane bend being the most intense
at 387 \kmol. Compared to the anti-symmetric stretch of H$_2$O at 70
\kmol\ and CO$_2$ at $\sim$475 \kmol, which are both considered to be
intense transitions, the present $\nu_2$ intensity should be
considered intense enough for detection \textit{via} IR
spectroscopy even at low concentrations. Additionally, the other vibrational 
fundamentals of \alane\ with
intensities of 251 and 234 \kmol\ are also substantially greater than
the aforementioned stretches in water and carbon dioxide. Looking at the
spectral profile, $\nu_4$ sits at 12.67 $\mu$m putting it around the
13 $\mu$m spectral feature that is typically used as an
identifier for \alumina. While this does not really question
\alumina's presence based on this spectral feature, it does posit the
existence of another carrier of said spectral feature in CSM or the
ISM that may be attributed to the present aluminum nitride
system. Nevertheless, the computed vibrational data herein should be
especially important for the detection of \alane\ that would be supportive
for confirming the reaction pathway of formation proposed in this work
and from previous computational reaction pathways \citep{Grosselin22, Flint23b}.

\begin{table}[ht]
\centering
\caption{Vibrational frequencies (\cm), IR intensities (km
  mol$^{-1}$) with absorption cross sections in parentheses (10$^{-14}$ cm$^2$),
  and wavelength ($\mu$m) for C$_{3v}$ AlNH$_6$}
\label{alnh6_freq}
\begin{widetabular}{\textwidth}{llcrrrr}
Mode        & Symm. & Desc.            & Harm.  & Anharm. & $\mathnormal{f}$ ($\sigma$) & $ \lambda$ \\
\midrule
 $\nu_{1}$  & $E$   & NH Stretch       & 3588.9 & 3411.8  & 45 (5) & 2.93 \\
 $\nu_{2}$  & $A_1$ & Sym. NH Stretch  & 3468.8 & 3322.4  & 16 (2) & 3.01 \\
 $\nu_{3}$  & $A_1$ & Sym. AlH Stretch & 1885.7 & 1819.2  & 36 (6) & 5.50 \\
 $\nu_{4}$  & $E$   & AlH Stretch      & 1865.1 & 1800.4  & 331 (50) & 5.55 \\
 $\nu_{5}$  & $E$   & NH2 Rocking      & 1665.3 & 1624.4  & 19 (3) & 6.16 \\
 $\nu_{6}$  & $A_1$ & HNAl Sym. Bend   & 1269.0 & 1207.8  & 135 (30) & 8.28 \\
 $\nu_{7}$  & $E$   & HAlN Rocking     & 790.8  & 779.1   & 271 (100) & 12.84 \\
 $\nu_{8}$  & $A_1$ & HAlN Sym. Bend   & 773.9  & 755.7   & 515 (200) & 13.23 \\
 $\nu_{9}$  & $E$   & OPB              & 706.8  & 666.1   & 15 (10) & 15.01 \\
 $\nu_{10}$ & $A_1$ & AlN Stretch      & 426.4  & 389.3   & 11 (20) & 25.69 \\
 $\nu_{11}$ & $E$   & OPB              & 372.6  & 352.2   & 2 (-) & 28.40 \\
 $\nu_{12}$ & $A_2$ & Torsion          & 119.4  & 21.4    & 1 (-) & 467.29 \\
 ZPT &          &  & &12724.7 \\
\midrule
\end{widetabular}
\end{table}

The first intermediate, AlNH$_6$, exhibits the most notable
vibrational transition intensities of the two formed structures
studied herein, shown in Table \ref{alnh6_freq}. AlNH$_6$'s $\nu_8$
frequency, the H$-$Al$-$N symmetric bending motion, at 515 \kmol is 30
\kmol\ larger than the CO$_2$ motion mentioned above. Further,
AlNH$_6$ $\nu_4$, $\nu_6$, and $\nu_7$ exhibit transition intensities
of 331, 135, and 271 \kmol, respectively. While AlNH$_6$ exhibits
multiple intense vibrational transitions, due to the nature of its
synthesis, it may be a short-lived species as it will likely progress
to AlNH$_4$ or regress back to the reactants. However, even with a
relatively small concentration, its intense transitions may still be
seen in the IR. Rotationally, AlNH$_6$ also exhibits the largest
permanent dipole moment of the three molecules studied, at 5.40
D. Compared to the previously detected \citep{Tenenbaum10} AlOH molecule with a 
permanent dipole of 1.11 D \citep{Fortenberry20AlOH}, AlNH$_6$ may be a
suitable candidate for potential radioastronomical observation, but may be 
limited by shorter lifetimes as discussed above.
Regardless, like AlH$_3$, AlNH$_6$ also exhibits spectral features in
and around the 13 $\mu$m dust feature as well as the associated 11, 20
and 28 $\mu$m. Namely, the aformentioned $\nu_8$ frequency, with the 515
\kmol\ intensity, sits directly at 13.23 $\mu$m, as shown in Table
\ref{alnh6_freq}. It should be noted that the mid- to far-IR regions where 
these features are located are dominated by larger dust grains than present 
potential grain nucleating species. Regardless, a laboratory and observational 
study of AlNH$_6$ is warranted utilizing the novel reference data provided
herein in order to assist in the astrophysical detection of the first
intermediate along an alternative pathway for formation for dust
grains in CSM or the ISM.

\begin{table}[ht]
\caption{Vibrational frequencies (\cm), IR intensities (km mol$^{-1}$) 
with absorption cross sections in parentheses (10$^{-14}$ cm$^2$), and 
wavelength ($\mu$m) for C$_{2v}$ AlNH$_4$}
\label{alnh4_freq}
\begin{widetabular}{\textwidth}{llcrrrrr}
 Mode       & Symm. & Desc. & Harm.  & Anharm. & $\mathnormal{f}$ ($\sigma$) & $\lambda$ & Prev.$^{a/b}$ \\
\midrule 
 $\nu_{1}$  & $b_2$ & Anti-symm. NH stretch  & 3673.6 & 3495.9  & 27 (3) & 2.86   & --/3496.8     \\
 $\nu_{2}$  & $a_1$ & Symm. NH Stretch       & 3582.7 & 3421.5  & 31 (4) & 2.92   & 3499.7/3421.7 \\
 $\nu_{3}$  & $b_2$ & Anti-symm. AlH stretch & 1961.2 & 1895.3  & 257 (40) & 5.28   & 1899.3/1895.5 \\
 $\nu_{4}$  & $a_1$ & Symm. AlH Stretch      & 1957.5 & 1893.5  & 81 (10) & 5.28   & 1891.0/1894.7 \\
 $\nu_{5}$  & $a_1$ & Symm. HNAl stretch     & 1582.0 & 1552.1  & 28 (5) & 6.44   & 1541.6/1552.7 \\
 $\nu_{6}$  & $a_1$ & AlN stretch            &  836.9 &  822.5  & 217 (90) & 12.16  & 818.7/822.7   \\
 $\nu_{7}$  & $a_1$ & Symm HAlN stretch      &  757.2 &  747.1  & 51 (20) & 13.39  & 755.0/747.9   \\
 $\nu_{8}$  & $b_2$ & Anti-symm. HNAl stretch&  732.6 &  717.5  & 145 (90) & 13.94  & 769.8/715.9   \\
 $\nu_{9}$  & $b_1$ & Al OPB                 &  614.4 &  606.0  & 165 (100) & 16.50  & 608.7/611.3   \\
 $\nu_{10}$ & $a_2$ & Anti-symm. Torsion     &  495.0 &  469.5  & 0 (0)  & 21.30 & --/465.8      \\
 $\nu_{11}$ & $b_1$ & N OPB                  &  448.9 &  442.9  & 236 (300) & 22.58  & 518.3/426.1   \\
 $\nu_{12}$ & $b_2$ & Anti-symm. HAlN stretch&  426.0 &  422.8  & 19 (30) & 23.65  & --/431.4      \\
 ZPT        &       &                        &        & 8427.6  &      &           &               \\
\end{widetabular}
\\
\raggedright
$^a$ Previous Ar matrix attributions \citep{Himmel00}.\\
$^b$ Previous F12-TZ anharmonic vibrational data \citep{Watrous21}.
\end{table}

The AlNH$_4$ molecule has both previous Ar matrix spectroscopic data \citep{Himmel00}
and theoretical vibrational frequency computations \citep{Watrous21}. The previous 
vibrational frequency studies are computed at the F12-TZ level of theory much 
like the rovibrational spectroscopic data provided herein. While the present 
computational study utilizes a normal coordinate system to compute the QFF procedure, 
the previous study utilizes a symmetry internal coordinate system that is comparable 
to the present normal coordinate system. Any deviation between the two methods should 
be considered an effect of the difference between the construction of the two 
coordinate systems. In any case, while, AlNH$_6$ contains more intense vibration 
transitions, AlNH$_4$ still exhibits exceptionally intense transitions. Shown in Table
\ref{alnh4_freq}, the AlNH$_4$ molecule's most intense transition is its
anti-symmetric Al$-$H stretching motion, $\nu_3$, of 257 \kmol. Like
AlNH$_6$, AlNH$_4$ also contains multiple intense vibrational
transitions such as $\nu_6$, $\nu_8$, $\nu_9$, and $\nu_{11}$ at 217, 145,
165, and 236 \kmol, respectively. Again, while most of the these
transitions are less intense than AlNH$_6$, the above intensities are
still relatively intense suggesting
AlNH$_4$ is a strong candidate for potential astrophysical detection
utilizing IR spectroscopy. Additionally, this species may be
longer-lived in its formation environment in CSM making it an even
stronger candidate for astronomical observational detection, perhaps, than
AlNH$_6$ itself.

In terms of its dipole moment, the AlNH$_4$ molecule's is much smaller at 
1.08 D. While AlNH$_4$ can still be observed rotationally given a 
high enough column density, with such a
small dipole moment AlNH$_4$ may be better suited for
detection via high resolution IR spectroscopy which can be achieved
\textit{via} the JWST. Also, AlNH$_4$ contains multiple spectral
features that fall in or around the 13 $\mu$m feature and other associated
spectral lines. Specifically, $\nu_7$ and $\nu_8$ have wavelengths at 13.39 and
13.94 $\mu$m, respectively, shown in Table \ref{alnh4_freq}. Again,
this suggests that more than \alumina\ or the other associated
suspected carriers exhibit this spectral feature, but this does
suggest that species containing aluminum may contribute to the
features more than initially observed or speculated. To that end, the
novel vibrational spectroscopic data provided herein are necessary for
further laboratory and possible astronomical observational
investigations into the proposed formation pathway that will
potentially assist in characterizing another potential player in dust
grain formation.

\section{Conclusions}
The reaction pathway of \alane\ and \ammonia\ leads to the formation of
larger aluminum nitride molecular systems. This pathway must first
overcome a barrier of 1.9 \kcal\, which is fully achievable in high
temperature environments ($\sim$1000 K) such as warmer inner
protoplanetary disks and circumstellar envelopes of AGB stars, 
but not necessary to begin the process of formation. The rest 
of the pathway is submerged compared to the reactants and is
available given the latent energy present in the molecular system. The
final step of the presently studied pathway involves the formation of
a four-membered cyclic ring that is also the only step on the pathway
to have a submerged product over its preceding intermediate. Even so,
the present reaction pathway provides a novel, potential formation
mechanism for Al-bearing dust grains containing nitrogen as
suggested in the studies of the Murchison CM2 chondritic meteorite and 
gives a hint at where such nucleation can take place in protoplanetary disks.

The AlNH$_4$ molecule investigated herein is the most likely candidate for
potential astronomical observation \textit{via} current rovibrational 
spectroscopic technologies in order to support this proposed reaction
pathway. AlNH$_4$ contains multiple vibrational transitions with
exceptionally large intensities, notably the 1895.3 \cm\ frequency with
the 257 \kmol\ intensity and the 442.9 \cm\ frequency with the 236 \kmol\
intensity. While AlNH$_4$ has a smaller dipole moment of 1.08 D, if there
is sufficient column density, this species should still be observable
utilizing current radioastronomical telescopes. The rovibrational analysis
of AlNH$_6$ suggests that the most intense vibrational transition studied
here is the 515 \kmol\ intensity corresponding to the 755.7 \cm\
frequency. Additionally, the 5.40 D dipole moment feature calculated is
the largest of the three studied. However, given the AlNH$_6$ molecule's
likely shorter timescales within the reaction pathway, it may not be a suitable
candidate for rovibrational detection in CSM or the ISM. The AlH$_3$
molecule also exhibits exceptionally intense vibrational transitions,
especially the 707.8 \cm\ frequency with an intensity of 387 \kmol. Like
the AlNH$_6$ molecule, the \alane\ molecule may react too quickly, with
ammonia, water, or other circumstellar or interstellar denizens before it can be
detected. However, the spectral features give it a strong chance of observation 
with \textit{JWST}. That being said, while all three species contain strong IR
features and some strong rotational features, AlNH$_4$ may be the most
likely candidate for potential astronomical observation.

The three Al-containing molecules studied in the present work contain
vibrational frequencies that fall in or around the 13 $\mu$m spectral
feature attributed to Al$-$O class of molecules. Additionally, the
computed vibrational profile shows the presence of vibrational frequencies
at the associated 11, 20, and 28 $\mu$m carrier features. The spectral
features found in the present Al-containing species imply that dust
containing some form of aluminum oxide may not be the only source of the
13 $\mu$m dust feature, and the aluminum nitride species studied herein may
provide other alternative answers to the question of the origins of
such features. These computed vibrational analyses warrant further
laboratory, theoretical, and observational investigations into the 
present reaction profile and other,
higher-order aluminum nitride clusters. These investigations will
assist in the search for Al-containing species that may be present in
CSM and the ISM, but have been uncharacterized and understudied in the
present literature.

\section{acknowledgment}
This work is supported by NASA Grants NNH22ZHA004C and
22-A22ISFM-0009 and by the University of Mississippi's College of
Liberal Arts. The computational support is from the Mississippi
Center for Supercomputing Research funded in part by NSF Grant
OIA-1757220. CZP would like to thank Dr.~Brent R. Westbrook for his 
revision of this manuscript. The authors would like to thank the reviewer 
for their important insight in the introduction and discussion of this 
manuscript.
 
\bibliography{refs}{}
\bibliographystyle{aasjournal}

\end{document}